\documentclass[prx,twocolumn, superscriptaddress,amsmath,amssymb,longbibliography]{revtex4-1}

\usepackage{graphicx}
\usepackage{dcolumn}
\usepackage{bm}
\usepackage{lineno}
\usepackage[colorinlistoftodos]{todonotes}
\begin{document}
    
\title{{Nature of excitons and their ligand-mediated delocalization in nickel dihalide charge-transfer insulators}}

\affiliation{Department of Physics, Massachusetts Institute of Technology, Cambridge, Massachusetts 02139, USA}
\affiliation{Institut de Minéralogie, de Physique des Matériaux et de Cosmochimi, Sorbonne Université, CNRS UMR 7590,  4 Place Jussieu, 75005 Paris, France}
\affiliation{Materials Science and Engineering, School for Engineering of Matter, Transport and Energy, Arizona State University, Tempe, Arizona 85287, USA}
\affiliation{National Synchrotron Light Source II, Brookhaven National Laboratory, Upton, New York 11973, USA}
\affiliation{Debye Institute for Nanomaterials Science, Utrecht University, 3584 CG Utrecht, Netherlands}
\affiliation{These authors contributed equally to this work.}

\author{Connor A. Occhialini}
\email{caocchia@mit.edu}
\affiliation{Department of Physics, Massachusetts Institute of Technology, Cambridge, Massachusetts 02139, USA}
\affiliation{These authors contributed equally to this work.}

\author{Yi Tseng} 
\email{tsengy@mit.edu}
\affiliation{Department of Physics, Massachusetts Institute of Technology, Cambridge, Massachusetts 02139, USA}
\affiliation{These authors contributed equally to this work.}

\author{Hebatalla Elnaggar} 
\affiliation{Institut de Minéralogie, de Physique des Matériaux et de Cosmochimi, Sorbonne Université, CNRS UMR 7590,  4 Place Jussieu, 75005 Paris, France}

\author{Qian Song} 
\affiliation{Department of Physics, Massachusetts Institute of Technology, Cambridge, Massachusetts 02139, USA}

\author{Mark Blei} 
\affiliation{Materials Science and Engineering, School for Engineering of Matter, Transport and Energy, Arizona State University, Tempe, Arizona 85287, USA}

\author{Seth Ariel Tongay} 
\affiliation{Materials Science and Engineering, School for Engineering of Matter, Transport and Energy, Arizona State University, Tempe, Arizona 85287, USA}

\author{Valentina Bisogni}
\affiliation{National Synchrotron Light Source II, Brookhaven National Laboratory, Upton, New York 11973, USA}

\author{Frank M. F. de Groot} 
\affiliation{Debye Institute for Nanomaterials Science, Utrecht University, 3584 CG Utrecht, Netherlands}

\author{Jonathan Pelliciari} 
\affiliation{National Synchrotron Light Source II, Brookhaven National Laboratory, Upton, New York 11973, USA}

\author{Riccardo Comin}
\email{rcomin@mit.edu}
\affiliation{Department of Physics, Massachusetts Institute of Technology, Cambridge, Massachusetts 02139, USA}

\date{\today}

\begin{abstract}
{

The fundamental optical excitations of correlated transition-metal compounds are typically identified with multielectronic transitions localized at the transition-metal site, such as $dd$ transitions. In this vein, intense interest has surrounded the appearance of sharp, below band-gap optical transitions, i.e. excitons, within the magnetic phase of correlated Ni$^{2+}$ van der Waals magnets. The interplay of magnetic and charge-transfer insulating ground states in Ni$^{2+}$ systems raises intriguing questions on the roles of long-range magnetic order and of metal-ligand charge transfer in the exciton nature, which inspired microscopic descriptions beyond typical $dd$ excitations. Here we study the impact of charge-transfer and magnetic order on the excitation spectrum of the nickel dihalides (NiX$_2$, X $=$ Cl, Br, and I) using Ni-$L_3$ resonant inelastic x-ray scattering (RIXS). In all compounds, we detect sharp excitations, analogous to the recently reported excitons, and assign them to spin-singlet multiplets of octahedrally-coordinated Ni$^{2+}$ stabilized by intra-atomic Hund's exchange. Additionally, we demonstrate that these excitons are dispersive using momentum resolved RIXS. Our data evidence a ligand-mediated multiplet dispersion, which is tuned by the charge-transfer gap and independent of the presence of long-range magnetic order. This reveals the mechanisms governing non-local interactions of on-site $dd$ excitations with the surrounding crystal/magnetic structure, in analogy to ground state superexchange. These measurements thus establish the roles of magnetic order, self-doped ligand holes, and intersite coupling mechanisms for the properties of $dd$ excitations in charge-transfer insulators.}
\end{abstract}

\maketitle

%Introduction
{The recent demonstration of magnetic order in correlated transition-metal van der Waals (vdW) materials to the ultrathin limit has led to an increased interest in their excitonic responses and coupling to magnetism. In contrast to uncorrelated, direct bandgap semiconductors exhibiting Wannier-type interband excitons \cite{Knox1983}, the below bandgap excitations of strongly correlated transition-metal compounds are typically interpreted in terms of localized transitions between distinct spin/orbital configurations of the transition-metal ions. Also known as $dd$ or ligand-field transitions \cite{ballhausen1962}, such excitations may be equivalently described as Frenkel-type excitons \cite{Agranovich1968}. Of particular interest is the utility of $dd$ excitation optical responses for measuring and tuning magnetic states, as exemplified by the observation of helical ligand-field luminescence in ferromagnetic Cr trihalides \cite{Seyler2018, Zhang2019}, the linearly polarized absorption/emission from excitons in the Ni$^{2+}$ vdW magnets NiI$_2$ \cite{Son2022} and NiPS$_3$ \cite{Kang2020a, Wang2021, Kim2023, Ergecen2022, Jana2023}, and associated photo-induced magnetic properties \cite{Belvin2021,Afanasiev2021, Zhang2022}. Clarifying the microscopic origin of such excitonic states, particularly their coupling mechanism to the local spin degree of freedom and long-range magnetism, is essential for continued progress towards functional applications and for the optical characterization of magnetic ground states in vdW materials.}

{Here, we focus on the triangular-lattice nickel dihalide antiferromagnets NiX$_2$ (X = Cl, Br, I) based on Ni$^{2+}$ ions ($3d^8$) to study their multiplet spectra versus ligand, temperature and momentum using Ni-$L_3$ edge resonant inelastic X-ray scattering (RIXS). The ligand-field spectra of Ni$^{2+}$ systems have been the subject of intensive study in classical optical literature \cite{Kozielski1972, Day1976, Kuindersma1981a, Robbins1976, Pollini1980, Giordano1978, Benedek1979} and more recent studies \cite{Kang2020a, Son2022, Afanasiev2021, Belvin2021, Ergecen2022, Jana2023, Wang2021, Kim2023, Wang2022}. Specifically, several recent investigations report the emergence of sharp excitons below the magnetic transition temperatures of correlated Ni$^{2+}$ vdW magnets \cite{Kang2020a, Son2022, Ergecen2022, Jana2023, Kim2023, Wang2021}} These excitons were associated to spin-entangled Zhang-Rice triplet-to-singlet excitations stabilized by long-range magnetic order and magnetic coherence \cite{Kang2020a, Son2022}. The Zhang-Rice mechanism is motivated by the charge-transfer insulator nature of the {electronic} ground state in Ni$^{2+}$ systems \cite{Zaanen1986, VanDerLaan1986}. The electronic states are an admixture between local $3d^8$ and $3d^9\underline{L}$ configurations, where $3d^9\underline{L}$ represents self-doped ligand hole electronic configuration \cite{Zaanen1986, VanDerLaan1986, DeGroot2008}. From this configuration, analogs to Zhang-Rice states may arise as observed in doped copper oxides \cite{Khomskii2014, Zhang1988, Tjeng1997}. {Despite this, the reported excitons bear a strong resemblance to optically spin-forbidden multiplet transitions previously revealed by optical spectroscopy \cite{Day1976, Robbins1976, Kozielski1972, Banda1986, Kuindersma1981a, Rosseinsky1978, Pollini1980, Giordano1978, Benedek1979}. Based on this dichotomy, the proposed Zhang-Rice mechanism for stabilizing these exciton states and the role of long-range magnetic order requires further scrutiny.}

{The nickel halides provide a platform to assess each of these aspects directly. First, the nickel dihalides are vdW magnets exhibiting distinct ligand-tuned magnetic ground states, ranging from $C$-type antiferromagnetic (AFM) to non-collinear spin structures \cite{Lindgard1975, Day1976, Regnault1982, Song2022, Son2022}. Furthermore, they constitute an archetypal series of charge transfer insulators with systematically tuned Ni-$X$ covalency and charge-transfer gap, $\Delta$, as previously revealed through both X-ray photoemission and X-ray absorption spectroscopy (XPS/XAS) \cite{VanDerLaan1986, Zaanen1986}. However, the impact that this strongly ligand-tuned $\Delta$ has on the ground state multiplet excitations has not been investigated in detail. As we show in this work, the simultaneous tuning of magnetic order and self-doped ligand holes through the charge-transfer gap establishes their roles in the emergence of the exciton states, and in their fundamental parameters (namely dispersion, microscopic nature, and temperature effects). While $dd$ excitations are nominally dipole forbidden in optics, RIXS provides direct spin- and dipole-allowed sensitivity to the zero-boson multiplet excitations \cite{DeGroot2008}. Such measurements are thus crucial to unravel their microscopic nature, and intrinsic evolution with both temperature and momentum to inform a proper interpretation of their manifestation in optical experiments.}

{From our RIXS measurements, we observe sharp (nearly resolution-limited) excitonic peaks in all NiX$_2$ (X $=$ Cl, Br, I) compounds, confirming their universality in Ni$^{2+}$ charge-transfer insulators. The ubiquity of these excitons stems from their microscopic nature, which we assign as spin-singlet $(S = 0)$ multiplet ($dd$) excitations of ${}^1A_{1g}$/$^1E_g$ symmetry \footnote{Throughout the paper, we will use the terms exciton/multiplet/$dd$-excitation interchangeably. Such terminology is consistent with the usage of ``Frenkel excitons'' in reference to multiplet transitions in the early optical literature \cite{Sell1967, Agranovich1968, Kozielski1972, Banda1986, Freeman1967, Freeman1968, Kuindersma1981a, Robbins1976}. However, referring to these states as multiplet transitions avoids confusion with excitons of distinct microscopic origin (e.g. Wannier excitons) routinely observed in semiconductors.}. These features are characteristic of Ni$^{2+}$ ions in octahedral symmetry, and are broadly consistent with their original identification through optical spectroscopy \cite{Day1976,Robbins1976, Kozielski1972, Banda1986, Kuindersma1981a, Rosseinsky1978, Pollini1980, Giordano1978, Benedek1979}}. Crucially, while the energies of these singlet excited states are strongly affected by the ligand and their ionic character, the existence of these excitons does not necessitate particular charge-transfer contributions or long-range magnetic order \cite{Nag2020, DeGroot2008}. Using charge-transfer multiplet (CTM) theory, we show the strong ligand-dependence results from {an effective screening of the intra-atomic Hund's exchange interaction (e.g. nephelauxetic effect)} \cite{DeGroot2008, Zaanen1986, Brik2016, Okada1992, Kitzmann2022} due to the increasing contribution of self-doped ligand hole ($3d^9\underline{L}$) states at small $\Delta$. This systematically establishes the primary effects of the metal-ligand hybridization/covalency, {and provides a complete description of the multiplet structure in the highly ionic (NiCl$_2$) and strongly covalent (NiI$_2$) limits. This evidence resolves key discrepancies in the peak assignments and their microscopic interpretations found throughout the literature.}

{Additionally, we uncover a finite momentum dispersion of the spin forbidden ${}^1A_{1g}/{}^1E_g$ multiplets.} As the charge-transfer energy is reduced by changing the halogen ligand, their dispersive bandwidth increases. This dispersion is independent from the magnetic dynamics, demonstrated through a direct comparison to the spin-excitation dispersions. Furthermore, the excitons and their dispersive behavior persist far above the magnetic ordering temperatures, demonstrating an exciton delocalization regardless of the presence of long-range magnetism. To explain this, we propose a simple charge-transfer induced exciton delocalization mechanism, determined from the dominant ligand-mediated orbital hopping pathways on the 2D triangular lattice. This effect can be viewed as a natural consequence of the increased metal-ligand hybridization in a crystalline environment. Our results demonstrate that the microscopic interactions stabilizing these excitons are the multielectron interactions at the nickel site and its local symmetry. This demonstrates how their fundamental energies and degree of delocalization may be tuned through the charge transfer gap. 

{ Our paper is organized as follows: (i) We present the experimental Ni-$L_3$ edge RIXS and XAS data versus ligand in the nickel dihalide series in Section I.A. (ii) We model the spectra with NiX$_6$ cluster calculations using charge-transfer multiplet theory to establish the microscopic origin of each excitation, and determine the role of the self-doped ligand holes in the ground and excited states in Section I.B. (iii) We present momentum and temperature dependence of the spin-forbidden ${}^1A_{1g}/{}^1E_g$ multiplets in comparison with spin-excitation dispersions in Section I.C, and (iv) we propose a microscopic model for the multiplet dispersion in Section I.D. Finally, we discuss the implications of these results in Section II and conclude in Section III.}

%Figure 1
\begin{figure*}[htb!]
\centering
\includegraphics[width=\textwidth]{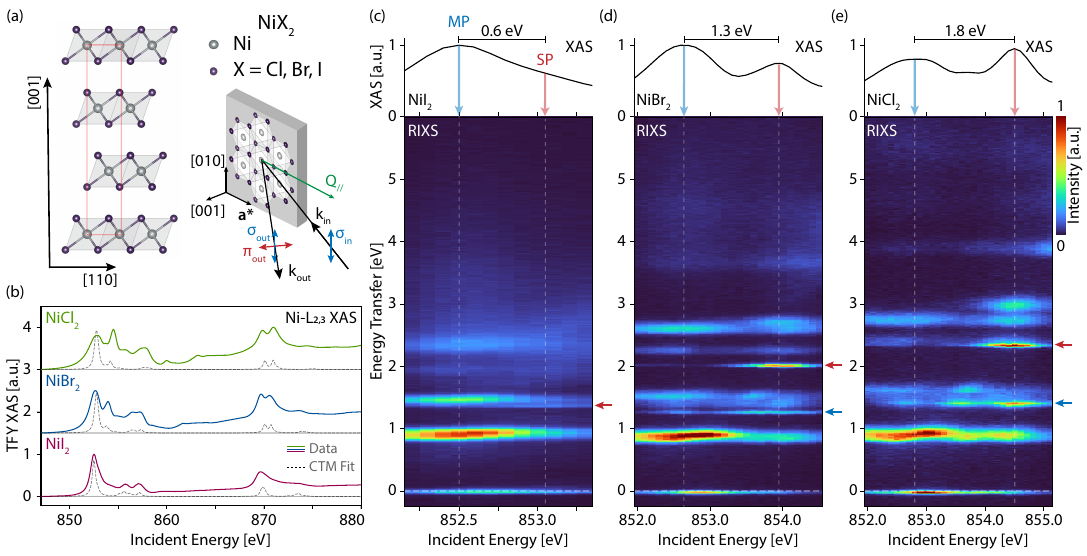}
\caption{(a), The layered rhombohedral (space group $R\bar{3}m$) structure of NiX$_2$ dihalide compounds, highlighting the triangular lattice of magnetic Ni$^{2+}$ ions and the RIXS scattering geometry at grazing incidence (see text). (b), Ligand-dependent Ni-$L_3$/$L_2$ edge XAS spectra at $T = 40$ K. {Corresponding XAS fits from charge transfer multiplet calculations are the dashed grey lines (see text).} (c-e), Incident-energy dependent RIXS maps for each compound across the main and side peaks at the Ni-$L_3$ edge (bottom) with the corresponding TFY XAS spectra (top). The XAS spectra indicate the main/side peak (MP/SP) XAS resonances with blue and red arrows, respectively, along with the MP-SP incident energy splitting for each compound. Red/blue arrows on the right axis of each RIXS map indicate the ${}^1A_{1g}$ - ${}^1E_g$ spin-singlet multiplets, resonant at the XAS SP.}
\label{fig:fig1}
\end{figure*}

\section{Experimental Results}

We performed XAS and RIXS measurements on high-quality single crystals of NiX$_2$ compounds grown by chemical vapor transport (see Methods). The RIXS and XAS data were acquired at the 2-ID SIX beamline of the National Synchrotron Light Source II, Brookhaven National Laboratory \cite{Dvorak2016}. RIXS measurements were performed with an energy resolution of $\Delta E = 31$ meV at the Ni-$L_3$ edge and XAS was recorded in total fluorescence yield (TFY). The samples were aligned with the $\mathbf{a}^*$ reciprocal lattice direction aligned in the scattering plane [Fig. \ref{fig:fig1}(a)], with grazing incidence geometry, $\sigma$ incident polarization, and $T = 40$ K unless otherwise specified. The temperature was chosen to be below the magnetic phase transitions for each compound. 

\subsection{Evolution of RIXS/XAS Spectra versus Ligand}

We begin by discussing the evolution of the Ni $L_3$-edge RIXS and XAS spectra versus ligand (X) for NiX$_2$ (X $=$ Cl, Br, I). Figure \ref{fig:fig1}(b) shows XAS spectra across the Ni ${L}_3$/${L}_2$ edges for each compound. At the $L_3$ edge, a clear double-peaked structure is observed for all compounds corresponding to a ``main'' peak (MP) around 852.7 eV, followed at higher energy by a ``side'' peak (SP) at 854.5 eV, 854.0 eV, and 853.1 eV for $\mathrm{NiCl}_2$, $\mathrm{NiBr}_2$, and $\mathrm{NiI}_2$, respectively [see Fig. \ref{fig:fig1}(c-e), top]. We note the pronounced self-absorption effect in the TFY XAS spectra \cite{VanDerLaan1986, DeGroot1994, Carboni2005, Wang2020}, leading to suppression of main $L_3$ edge intensity (see Supplemental Material (SM) \cite{SuppRef}). The energy separating the MP and SP increases with the ionic character of the compound. The latter is directly linked to higher charge-transfer gaps, $\Delta$, as discussed below. Additionally, several broader peaks are observed at higher energies ($E_i \simeq 855-860$ eV) and associated to charge-transfer satellites \cite{DeGroot2008, VanDerLaan1986, Okada1992}. An overall similar qualitative behavior is observed at the $L_2$ edge.

We subsequently measure RIXS spectra versus incident energy across the MP and SP resonances at the Ni $L_3$ edge for each compound [Fig. \ref{fig:fig1}(c-e)]. For Ni$^{2+}$ in $O_h$ crystal field, the ground state electronic configuration is $^3A_{2g}$ with $t_{2g}^6e_g^2$ orbital occupation and $S = 1$ arrangement of the half-filled $e_g$ states \cite{ballhausen1962, DeGroot1990}. Around $\Delta E = 950$ meV energy transfer, we identify a predominant Raman-like excitation which is nearly independent of ligand. This excitation is connected to the fundamental $t_{2g} \to e_g$ spin preserving ($\Delta S = 0$) crystal field excitation ($^3T_{2g}$), suggesting a similar $O_h$ crystal field energy scale ($10Dq$) across the series \cite{Nag2020}. { The independence of this energy scale with ligand can be rationalized by the balance of charge-transfer and metal-ligand hybridization contributions, both of which affect the covalent crystal field splitting as captured by our CTM calculations and discussed below \cite{Ushakov2011, Scaramucci2015, Khomskii2014}.} At higher-energies ($\Delta E = 1-3$ eV), rich excitation profiles are resolved with a strong ligand-dependence. These peaks are linked to the multiplet structure of Ni$^{2+}$ in $O_h$ symmetry \cite{DeGroot2008, Nag2020,ballhausen1962, DeGroot1998}. We highlight the uniquely sharp excitations around 1.38, 2.04, and 2.37 eV in I/Br/Cl [red arrows, Fig. \ref{fig:fig1}(c-e)], respectively. Remarkably, these peaks are nearly resolution-limited and resonant near the SP. Additional sharp excitations at the SP near $\Delta E =$ 1.3/1.45 eV in Br/Cl (blue arrows) are also revealed.

%Figure 2
\begin{figure*}[htb!]
\centering
\includegraphics[width = \textwidth]{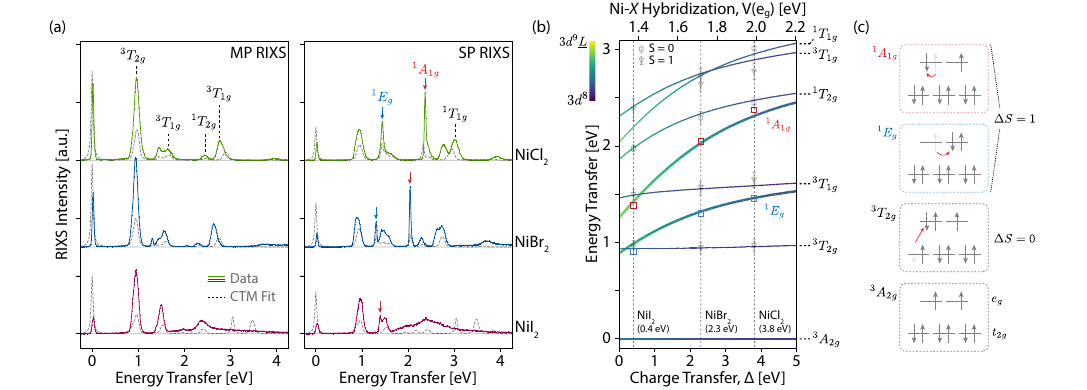}
\caption{(a), Ligand-dependent RIXS spectra at the MP (left) and SP (right) resonances. Intensity for each spectrum has been normalized by the total inelastic signal ($0.5$ $\to$ $5.0$ eV). The peaks are labelled with their corresponding multiplet term symbol in $O_h$ symmetry (see text). The sharp spin-singlet excitations (${}^1E_g$ and ${}^1A_{1g}$) are indicated for each compound with blue/red arrows, respectively. {Corresponding CTM calculations are shown as dashed grey lines overlaid with each experimental spectrum. The sharp features in the CTM calculations for NiI$_2$ above 3 eV are charge-transfer transitions, which are broad and overlap with fluorescence-like background in experiment (thus not well-resolved). The lower energy 3d$^8$ multiplets in all NiX$_2$ samples are sharp and well-captured by CTM calculations.} (b), Energy level diagram calculated from the CTM model as a function of charge transfer gap $\Delta$ and Ni-$X$ hybridization $V(e_g)$ in $O_h$ symmetry (spin-orbit coupling excluded for simplicity). {The ${}^1A_{1g}$ and ${}^1E_g$ term energies are highlighted with thick lines, and all calculated excitations are colored based on the $3d^8$/$3d^9\underline{L}$ character, as indicated by the color bar. Experimental energies for the ${}^1A_{1g}$/${}^1E_g$ peaks are shown as red/blue data points with other experimental multiplets indicated with triangles/squares for triplet/singlet terms, respectively.} Optimal fit values of $\Delta$ for each compound are indicated with vertical dashed lines. (c), Schematic representation of the low-energy $\Delta S = 0$/$\Delta S = 1$ $3d^8$ multiplet terms in $O_h$ symmetry.}
\label{fig:fig2}
\end{figure*}

The ligand-dependent RIXS spectra at the MP and SP resonances are summarized in Fig. \ref{fig:fig2}(a). The individual $dd$ transitions are assigned with term symbols in $O_h$ symmetry for NiCl$_2$ (top), based on our CTM calculations. Specifically, the higher/lower energy sharp peaks resonant at SP are ascribed to the spin flip $\Delta S = 1$ and ${}^1A_{1g}$/${}^1E_g$ multiplet terms, respectively \cite{Nag2020, DeGroot1998}, which preserve the ground state $t_{2g}^6 e_g^2$ orbital configuration. These spin-singlet multiplets are equivalent to the previously identified excitons in the optical regime, appearing at the same energies \cite{Kang2020a, Son2022, Kuindersma1981a, Banda1986, Day1976, Rosseinsky1978, Kozielski1972}. The salient features of the ligand dependence can be summarized by (i) a reduction of the MP-SP splitting in XAS, (ii) a reduction of multiplet energies that is most pronounced in the spin-singlet ${}^1 A_{1g}$ and ${}^1E_g$ excitations, and (iii) the resonant behavior of the $\Delta S = 1$ excitations at the ligand-dependent SP resonance.

\subsection{NiX$_6$ Cluster Calculations}

We next aim to quantitatively describe these ligand-dependent spectroscopic features and to provide a robust assignment of the electronic ground states and excitations. To do so, we employ CTM calculations as implemented in Quanty \cite{Haverkort2014,Haverkort2012,Lu2014}. The model reduces to a multielectronic calculation of a single NiX$_6$ cluster with $O_h$ symmetry, accounting for the Ni-$3d$ orbitals and the corresponding symmetrized ligand X-$np$ molecular orbitals \cite{Haverkort2014,DeGroot2008,VanDerLaan1986} (see Methods). { We restrict the present analysis to octahedral symmetry, while potential effects of the trigonal distortion are discussed in the Supplemental Material.} Figure \ref{fig:fig2}(b) shows the evolution of low-energy multiplets for the $3d^8 + 3d^9\underline{L} + 3d^{10} \underline{L}^2$ configuration as a function of $\Delta$. The evolution of the ${}^1 A_{1g}$ and the ${}^1 E_g$ excited states as a function of ligand charge-transfer are highlighted with thick lines [Fig. \ref{fig:fig2}(b)]. These excitations correspond to a nearly pure spin-flip $\Delta S = 1$ within the $|e_g\rangle$ manifold without transfer of orbital weight between the $t_{2g}$-$e_g$ states. Thus, they are stabilized from the ${}^3 A_{2g}$ ground state by the intra-atomic Hund's exchange \cite{ballhausen1962} (see also Ref. \footnote{We note the distinction between these local spin-state excitations (e.g. $S = 1$ $\to$ $S = 0$) and the single-/two-magnon excitations \cite{Li2022, Elnaggar2023, Nag2020}. The former, considered here, are $dd$ excitations stabilized by Hund's exchange. The latter are transitions of the spin projection $\Delta m_s = 1$ within the triplet ground state, with characteristic energies determined by the effective spin exchange \cite{DeGroot2008, ballhausen1962, DeGroot1998}.}). Their preservation of the $t_{2g}^6e_g^2$ ground state orbital configuration, in conjunction with their low-degeneracy, naturally accounts for their characteristically sharper linewidths compared to the other inter-configurational multiplets \cite{Freeman1967,Sell1967,Kozielski1972,Kitzmann2022}.

We identify the optimal parameters for each compound based on a minimal parameter fitting while keeping the Coulomb interactions at the nickel site fixed {(for a detailed description of the model and parameters, see SM \cite{SuppRef} and Methods).} {The optimized ligand-dependent CTM parameters $\Delta$ and the metal-ligand hybridization $V(e_g)$ are summarized in Table \ref{tab:tab1} and indicated as vertical dashed lines in Fig. \ref{fig:fig2}(b), with the experimental values of multiplet energies overlaid}. {The refined parameters are broadly consistent with previous reports from XPS and XAS \cite{VanDerLaan1986, Zaanen1986}, while our calculations are further restricted by the multiplet spectra which more accurately reflect the ground state Hamiltonian.} The simulated XAS/RIXS spectra determined from these parameters are shown as grey lines on top of the experimental data in Figs. \ref{fig:fig1}(c) and \ref{fig:fig2}(a). They reveal a good agreement with all salient features of the ligand dependence{, including the peak energies [Fig. \ref{fig:fig2}(b)], their resonance behavior/relative intensities [Fig. \ref{fig:fig2}(a)], as well as the MP-SP and charge transfer satellite structures in the XAS (Fig. \ref{fig:fig1}(b), for calculated RIXS maps to compare to experiments in Fig. \ref{fig:fig1}(c-e), see Supp. Fig. 14).}

\begin{table}[h]
\centering
\begin{tabular}{c||c|c|c|c}
 Ligand ($X$) & $\Delta$ [eV] & $V(e_g)$ [eV] & $\beta_{\text{eff}}$ & $J_H^\text{eff}$ [eV]\\
 \hline\hline
 Cl &  3.80 & 1.99 & 0.75 & 0.850\\

 Br &  2.30 & 1.72 & 0.64 & 0.722\\

 I  &  0.40 & 1.37 & 0.44 & 0.496\\
\end{tabular}
\caption{Ligand-dependence of $\Delta$ and $V(e_g)$ from CTM calculations. Coulomb interactions are fixed to atomic values and ionic contribution to $10Dq = 0.55$ eV is fixed for all ligands. The on-site Coulomb repulsions are fixed to $U_{dd} = 5.0$ eV and $U_{pd} = 7.0$ eV from photoemission experiments \cite{Zaanen1986}. Also shown is the phenomenological, effective ground state nephelauxetic reduction ($\beta_\text{eff}$) and intra-atomic Hund's exchange ($J_H^\text{eff}$) determined from corresponding ionic calculations in the Supplemental Material \cite{SuppRef}.}
\label{tab:tab1}
\end{table}

A consequence of reduced $\Delta$ is a larger mixing of the $3d^9 \underline{L}$ configuration into the ground and excited state $3d^8$ multiplets \cite{DeGroot2008, Okada1992, VanDerLaan1986, Son2022, Kang2020a, Nag2020, Takubo2007, Bisogni2016}. The energy level diagram in Fig. \ref{fig:fig2}(b) shows the evolution between $3d^8$ and $3d^9\underline{L}$ character resolved to each excitation. The ligand-hole character is excitation-dependent, with higher-energy excitations within a given orbital configuration (e.g. $t_{2g}^6 e_g^2$ vs. $t_{2g}^5 e_g^3$) displaying larger ligand character at a given $\Delta$, with the $|3d^9\underline{L}\rangle$ weight roughly commensurate to the energetic renormalization of each excitation. A similar situation determines the energy-dependent ligand-hole character and MP-SP reduction in the XAS intermediate states. Besides this energetic renormalization, all excitations remain direct analogs of the corresponding $3d^8$ multiplets as they stem only from the electronic configuration and point group symmetry \cite{SuppRef}. 

We conclude that the dominant role of ligand-hole states at the level of a single NiX$_6$ cluster is a renormalization of the intra-atomic Coulomb interactions in both the initial and final RIXS states. This is induced by delocalization of electronic density onto the ligand states (e.g. {nephelauxetic} effect \cite{DeGroot2008, Zaanen1986, Brik2016, Okada1992, Kitzmann2022}). This screening effect captures the evolution of the sharp singlet excitations, their resonance behavior, and the MP-SP evolution, which can be mapped to properties of Ni$^{2+}$ ions in $O_h$ symmetry without invoking emergent properties from the $3d^9\underline{L}$ configuration. From this assessment, the relevance of the recently proposed Zhang-Rice mechanism for these excitations can be ruled out \cite{Jana2023}. Thus, the sharp ${}^1 A_{1g}$ and ${}^1 E_g$ peaks are Ni$^{2+}$ $dd$-excitations arising from multielectronic interactions of a $3d^8$ electronic configuration in $O_h$ crystal field. One may therefore expect such excitations to be ubiquitous in isoelectronic systems close to the ionic limit \cite{Kitzmann2022,Nag2020,Wojnar2020,Brik2016,Son2022,Kang2020a,Kozielski1972,DeGroot2008}. {These conclusions are directly supported by calculations restricted to the purely ionic limit with $3d^8$ configuration and excluding charge-transfer processes summarized in the Supplemental Information \cite{SuppRef}. Indeed, the effect of ligand holes on the multiplet excitations can be captured by an effective nephelauxetic effect ($\beta_\text{eff}$), as summarized in Table \ref{tab:tab1}.}

\subsection{Exciton Dispersion and Relation to Magnetism}

%Figure 3
\begin{figure*}[t]%[tbhp]
\centering
\includegraphics[width=18.0cm]{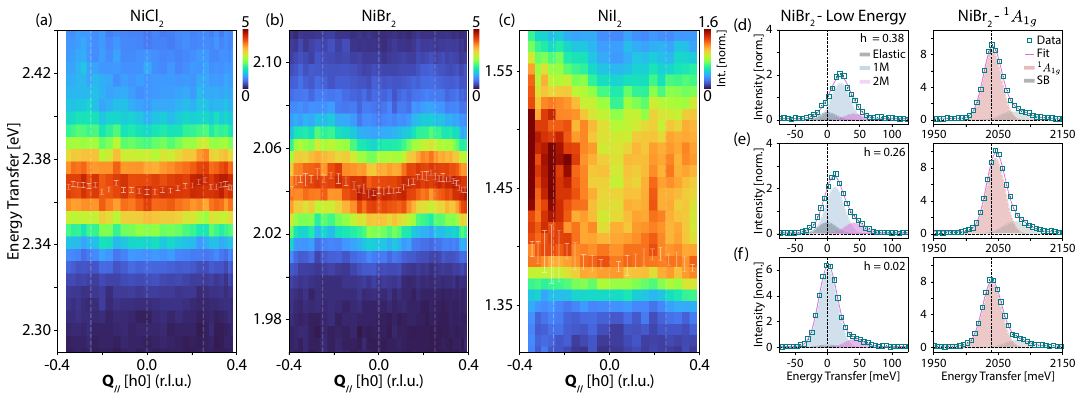}
\caption{Momentum-dependence of the ${}^1A_{1g}$ peak as a function of ligand at $T = 40$ K in panels (a, b, c) for Cl/Br/I respectively along the $\Gamma M$ direction. Momentum is reported as $\mathbf{Q_\parallel} = [h0]$ with $h$ expressed in reciprocal lattice units (r.l.u.). Color maps are normalized to the integrated intensity of the displayed region in (a, b) and to the region 1.35-1.42 eV in (c). Fitted points for the ${}^1A_{1g}$ peak are shown as overlaid white data points with error bars determined as standard errors from the fits. Figures (d-f) are example raw data along $\Gamma M$ with $h =$ 0.38, 0.26, and 0.02, respectively, for NiBr$_2$ both in the low-energy transfer (left) and ${}^1A_{1g}$ spectral regions (right), highlighting fits to the elastic (grey), single-/two-magnon (1M/2M, blue/purple, respectively) contributions, and the ${}^1A_{1g}$ peak (red) and its side band (SB, grey), with overall fit depicted in purple. Statistical error bars are indicated, which are smaller than the data points.}
\label{fig:fig3}
\end{figure*}

Having established the presence and properties of the exciton peaks in NiX$_2$ as a function of ligand, as well as their hybridized Ni/halogen nature, we now investigate their dependence on momentum and temperature to assess their delocalization beyond pure on-site $dd$ excitations \cite{Schlappa2012, Bisogni2015, Martinelli2023} and their connection to the magnetic order. 

We first report the dispersion of the ${}^1A_{1g}$ excitation for each ligand in Fig. \ref{fig:fig3}(a-c), measured at $T = 40$ K with momentum transfer along the $\mathbf{a}^*$ direction ($[h0]$ in reciprocal lattice units, r.l.u.) and with incident energy tuned to the SP resonance (Figs. \ref{fig:fig1} and \ref{fig:fig2}). We resolve an electronic dispersion with bandwidth $\delta E \simeq 3.4 \pm 1.2$ meV in $\mathrm{NiCl}_2$, $\simeq 8.2 \pm 1.3$ meV in $\mathrm{NiBr}_2$, and $\simeq 9.6 \pm 3.0$ meV in $\mathrm{NiI}_2$. Representative fits for spectra at selected momentum-transfer points for $\mathrm{NiBr}_2$ are shown in Fig. \ref{fig:fig3}(d-f) showing both the elastic/magnon and ${}^1A_{1g}$ spectral regions. Both spectral regions are from the same spectra at a given $\mathbf{Q}_\parallel$, recorded at the SP resonance where single- and two-magnon contributions are observed \cite{Nag2020, Li2022, Elnaggar2023, DeGroot1998}. {The spectra are relatively aligned using $\Delta S = 0$ multiplet excitations, which assumes these excitations are non-dispersive. This assumption can be justified by their high-multiplicity, significant phonon broadening, and the relatively low contribution of the $3d^9\underline{L}$ configuration, leading to the same spectral center of mass (i.e., lack of apparent dispersion). The corresponding analysis is discussed further in the Supplemental Material alongside additional discussion of polarization cross-section/multiplet fine structure effects \cite{SuppRef}.} Momentum-dependence for the ${}^1E_{g}$ excitation is also resolved in $\mathrm{NiCl}_2$ and $\mathrm{NiBr}_2$ \cite{SuppRef}.

\begin{figure*}[htbp!]
\centering
\includegraphics[width=18.0cm]{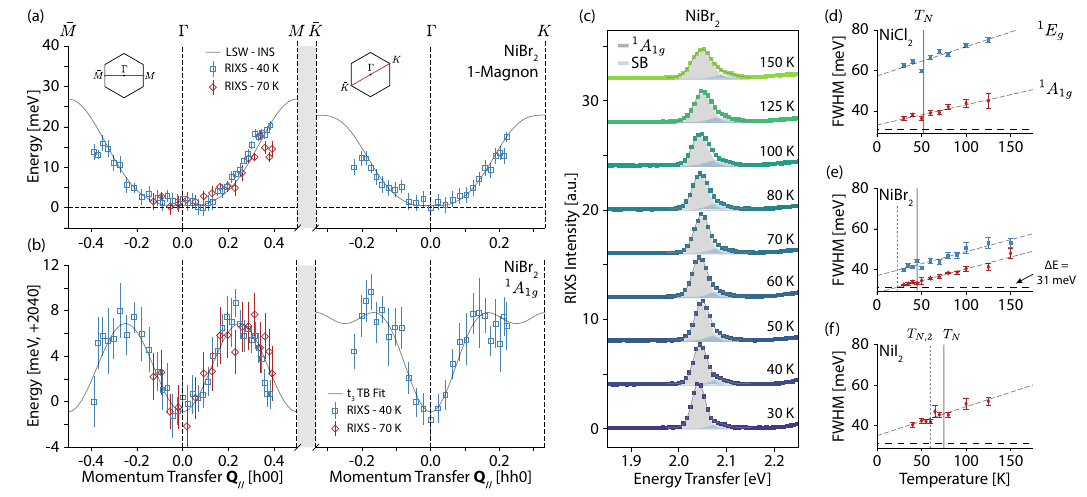}
\caption{Momentum-dependence of the single-magnon (a),  and the ${}^1A_{1g}$ excitation (b) in NiBr$_2$ along the $\Gamma M$ (left) and $\Gamma K$ (right) momentum-space cuts. Blue squares, $T = 40$ K and red diamonds, $T = 70$ K. Experimental single-magnon energies are compared to LSW theory, accounting for intra-layer exchange up to third nearest neighbor ($J_3$) using experimentally-determined parameters from Ref. \cite{Regnault1982}. The ${}^1A_{1g}$ data is fit along both $\Gamma M$ and $\Gamma K$ with a tight-binding (TB) model considering only third-nearest neighbor hopping ($t_3$). The LSW and TB curves are shown as solid grey lines in (a, b), respectively. (c), Representative temperature-dependence of the ${}^1A_{1g}$ excitation in NiBr$_2$ highlighting the temperature-dependent linewidth. Gaussian fits to the ${}^1A_{1g}$ and its side-band (SB, see text) are the filled grey/blue curves, respectively. Temperature-dependent linewidth of the ${}^1A_{1g}$ (red) and ${}^1E_g$ (blue) peaks for (d) NiCl$_2$, (e) NiBr$_2$ and (f) NiI$_2$. Linear fits (dashed lines) highlight a linear broadening of each peak with increasing temperature. Horizontal dashed lines denote the experimental resolution ($\Delta E = 31$ meV) for all measurements. Vertical solid lines indicate the $C$-type AFM transition temperature $T_N \simeq$ 52, 45, and 75 K for Cl/Br/I, respectively, and dashed lines indicate the non-collinear magnetic phases $T_{N,2} \simeq$ 22 and 60 K for Br/I, respectively \cite{Lindgard1975,Day1976,Kuindersma1981}.
}
\label{fig:fig4}
\end{figure*}

To elucidate the microscopic origin of the exciton dispersion and its reciprocal space structure in more detail, we perform momentum-dependent RIXS measurements across the magnetic phase transition and along different high-symmetry directions in reciprocal space, with a focus on NiBr$_2$. In Fig. \ref{fig:fig4}(a,b), we plot the fitted energy dispersion for the low-energy magnon and ${}^1A_{1g}$ mode in $\mathrm{NiBr}_2$, with comparison along the $\Gamma M$ ($[h0]$ r.l.u.) and $\Gamma K$ direction ($[hh]$ r.l.u.). The magnon and ${}^1A_{1g}$ exciton dispersion along $\Gamma M$ are further compared at 40 K and 70 K, representing the layered AFM and paramagnetic phase of $\mathrm{NiBr}_2$, respectively \cite{Regnault1982,Day1976,Kozielski1972}. From these data, we infer marginal differences in the ${}^1A_{1g}$ dispersion across the magnetic phase transition [Fig. \ref{fig:fig4}(b)], with the primary temperature effect being an overall broadened linewidth with increasing temperature (as discussed below) \cite{SuppRef}. This implies that the exciton dispersion is present regardless of long-range magnetic order, and therefore is likely not mediated by it. 

The magnon dispersions are compared with linear spin-wave (LSW) calculations ([Fig. \ref{fig:fig4}(a)]) based on inelastic neutron scattering (INS) in the layered AFM phase \cite{Regnault1982}, showing good quantitative agreement (for similar comparisons in NiCl$_2$ \cite{Lindgard1975} see SM \cite{SuppRef}). Importantly, the magnon and the ${}^1A_{1g}$ peaks have qualitatively distinct dispersions along $\Gamma K$ and $\Gamma M$. {From Fig. \ref{fig:fig3}(a-c), we also note the qualitatively similar functional form of the ${}^1 A_{1g}$ dispersion across all compounds, which is independent of the disparate magnetic structures or spin excitation dispersions (see also Supp. Fig. S10).} These aspects support our assignment of a genuine dispersion of the ${}^1A_{1g}$ excitations. To quantify this, we construct a minimal tight-binding (TB) model based on isotropic hopping parameters $t_n$ up to $n = 3$ nearest-neighbor (NN) \cite{Cudazzo2015,Gombar2018,Agranovich1968,Freeman1967}. We find that the ${}^1A_{1g}$ dispersion is well-described by considering only the third-NN contribution, with the single parameter ($t_3$) fit for NiBr$_2$ shown in Fig. \ref{fig:fig4}(b). We note that the spin-excitations persist above the long-range ordering temperatures [Fig. \ref{fig:fig4}(a)], suggesting the presence of short-range magnetic correlations persisting to high-temperatures \cite{Nag2022, Braicovich2009, Braicovich2010}. An effect of these short range magnetic correlations for determining the spin-singlet multiplet dispersion cannot be ruled out directly, although we will argue for a more natural mechanism as evidenced by the ligand dependence (discussed below).

To further underscore the independence of the $\Delta S = 1$ multiplets from the magnetic order, we measure the temperature dependence at fixed momentum transfer across the magnetic phase transitions in each compound as reported in Fig. \ref{fig:fig4}(c-f). A monotonic linewidth broadening is revealed for both the ${}^1A_{1g}$ and ${}^1E_{g}$ modes without any significant change of spectral profiles across the magnetic phase transition temperatures for each $\mathrm{NiX}_2$ compound [Fig. \ref{fig:fig4}(d-f)]. For NiBr$_2$ [Fig. \ref{fig:fig4}(c)], additional spectral weight at higher energies ($\sim$30-40 meV) above the ${}^1A_{1g}$ peak is apparent, which is attributed to two-phonon side bands, consistent with previous optical experiments \cite{Robbins1976, Kozielski1972, Day1976, Giordano1978, Pollini1980, Benedek1979}. Importantly, the linewidth and intensity of the singlet peaks are independent of the magnetic phase. In addition, we do not observe any clear correlation between the thermal broadening slope or extrapolated zero-temperature linewidth with the either the magnetic transition temperatures (as may be expected for magnetic coherence) or with the $|3d^9 \underline{L}\rangle$ character of the excitations [Fig. \ref{fig:fig4}(d-f)]. Instead, we attribute the thermally-activated broadening to a Franck-Condon phonon-coupling effect \cite{Kozielski1972, Lee2014, DeGroot1990}.

%Figure 4

{ Before moving on, we summarize the experimental observations and what they imply regarding the role of magnetism for these multiplet states. At the level of localized excitations, the charge transfer multiplet calculations identify the energy scale stabilizing the ${}^1A_{1g}$/${}^1E_g$ exciton states as the Hund's exchange. Therefore, their microscopic origin is independent of magnetic order but the relationship of these excitations to the local spin degree of freedom at the nickel site is clarified. At the next level detail, the independence of the fine structure of these excitations from magnetic order is corroborated by their temperature-dependencies at fixed momentum, revealing no anomalies in the linewidth or intensity across the long-range magnetic ordering temperatures. This establishes that the multiplet excitations are independent of long-range magnetic order, and sufficient to microscopically describe the excitons in NiX$_2$. 

Separately, we observe a finite dispersion of the ${}^1A_{1g}$/${}^1E_g$ excitations for all ligands, independent of the disparate spin structures of the different compounds and qualitatively distinct from the spin excitation dispersions. In NiBr$_2$, both the spin excitations and multiplet dispersions are observed to persist above the long-range ordering temperature. This allows the possibility for a role of short-range correlations in the multiplet dispersion. However, the totality of experimental evidence strongly suggests that at leading order, there is no direct effect of either short- or long-range magnetic order on the bare multiplet states, or in their dispersive character. This leads us to consider a mechanism of exciton delocalization that is independent of magnetism, but instead directly connected to increased contributions of the self-doped ligand holes with reduced $\Delta$. Even if magnetism would enter a more exhaustive description, our data directly evidence that strong emphasis must be placed on the self-doped ligand holes.}

\subsection{Origin of Exciton Dispersion}

\begin{figure}[t]
\centering
\includegraphics[width=\columnwidth]{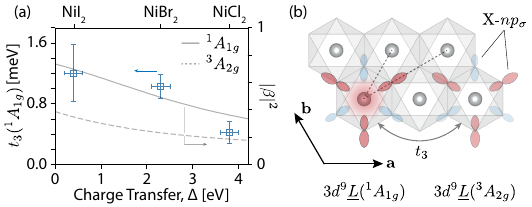}
\caption{(a), Ligand-dependence of the $t_3$ TB parameter from fits to the {\bf a}$^*$ dispersion data presented in Fig. \ref{fig:fig3} (a-c) (left axis, blue), along with the $3d^9\underline{L}$ character ($|\beta|^2$) of the excited state ${}^1A_{1g}$ (solid) and the ground state ${}^3A_{2g}$ (dashed) multiplets extracted from the CTM calculations in Fig. \ref{fig:fig2} (b) (right axis). (b), Schematic of the hopping pathways in the triangular lattice plane ($t_1$-$t_3$) and the proposed third-neighbor hopping mechanism between the $3d^9\underline{L}({}^1A_{1g})$ excited state and $3d^9\underline{L}({}^3A_{2g})$ ground state, mediated by ligand $p$-$p$ overlapping $\sigma$-bonding molecular orbitals.}
\label{fig:fig5}
\end{figure}

To examine the origin of the finite ${}^1A_{1g}$ dispersion, we consider the ligand-dependence of the dispersive bandwidth ($t_3$). This is displayed in Fig. \ref{fig:fig5}(a), extracted from the $\mathbf{a}^*$ dispersion in Fig. \ref{fig:fig3}(a-c), revealing an increase of the bandwidth with decreasing $\Delta$. {The ligand-dependent bandwidth follows the trend of the projected $3d^9\underline{L}$ character of the ${}^1A_{1g}$ state [$|\beta^2|$ in Fig. \ref{fig:fig5}(a)], implicating a ligand-mediated delocalization mechanism. This is also suggested by the dominance of third-NN interactions evidenced by the functional form of the dispersion, suggesting long-range interactions beyond direct $d\to d$ overlap. We interpret these features in analogy to the evolution of magnetic exchange interactions throughout the dihalide series, which have been analyzed in detail in the literature \cite{Lindgard1975, Regnault1982, Kuindersma1981, Amoroso2020, Song2022}.} The spin exchange is dominated by superexchange \cite{Botana2019, Takubo2007}, with moderate ferromagnetic $J_1$, negligible $J_2$ and antiferromagnetic $J_3$ parameters restricted to the 2D triangular lattice plane where $J_n$ is $n^\text{th}$ Ni-Ni neighbor exchange \cite{Amoroso2020}. While $J_1$ exhibits moderate stoichiometric dependence, $J_3$ is strongly ligand-dependent and is responsible for the difference in magnetic ground states across the series, including the non-collinear magnetic states in X = Br, I \cite{Amoroso2020, Regnault1982, Song2022, Rastelli1979}. Specifically, $J_3$ is mediated by superexchange pathways involving ligand $X$-$np$ molecular orbitals with large $\sigma$-type overlap \cite{Takubo2007}, as depicted in Fig. \ref{fig:fig5}(b).

From this picture, a dominance of the $t_3$ TB component in the ${}^1A_{1g}$ dispersion can be explained by the stronger $pp\sigma$ ligand-ligand transfer integrals between third-NN $3d^9\underline{L}$ $e_g$-symmetry molecular orbitals, in conjunction with the lowering of the charge transfer gap which mediates $pd$ electron transfer and increases the self-doped $3d^9\underline{L}$ character \cite{Takubo2007, DeGroot2008, Zaanen1986, Bisogni2016}. While superexchange occurs between Ni atoms in the ${}^3A_{2g}$ ground state, the dispersion we observe originates from interactions between an excited ${}^1A_{1g}$ impurity and a surrounding bath of ${}^3A_{2g}$ \cite{Freeman1967, Sell1967} [Fig. \ref{fig:fig5}(b)]. {The enhancement of $3d^9\underline{L}$ character upon excitation of the effective ${}^1A_{1g}$ defect compared to the ${}^3A_{2g}$ ground state [Fig. \ref{fig:fig5}(a)] could contribute to these observations due to selective enhancement of the excited state third-NN interactions, which are more sensitive to the ligand-hole contribution compared to nearest neighbor interactions \cite{Takubo2007}.}

These considerations are not unique to the ${}^1A_{1g}$ state, as a finite dispersion was also resolved for the ${}^1E_g$ state (see SM \cite{SuppRef}). For $^1E_g$, the dispersion is weaker and with opposite sign relative to the ${}^1A_{1g}$ but of a similar qualitative (sinusoidal) form { with increased bandwidth from NiCl$_2$ to NiBr$_2$ (see Supp. Fig. S10).} This suggests the sensitivity of the proposed hopping processes to the relative spin-orbital character of the excited and ground states, again in analogue to Pauli-restricted virtual hopping processes leading to superexchange \cite{Khomskii2014, Takubo2007, Freeman1967}. We note that since the ${}^1E_g$ and ${}^1A_{1g}$ states are $S = 0$ (non-magnetic), this should be interpreted as a multiplet-dependent effective transfer integral that is independent of the relative alignment of the surrounding ${}^3A_{2g}$ ground state spins -- that is, independent of the long-range magnetic order. This scenario is consistent with the insensitivity of the spin-singlet multiplets and their dispersion to the magnetic transitions (Fig. \ref{fig:fig4}). 

{We stress a distinction between the role of magnetism on the multiplet state and its dispersion from the effects of dispersive excitons on the surrounding magnetic order. We have shown that the excitons and their dispersion are independent from magnetic order. Meanwhile, the dispersion reveals the microscopic pathway mediating non-local interactions of the locally excited $S = 0$ magnetic defect with the surrounding magnetic sites. This was previously elaborated for excited state exchange interactions and their relevance to exciton-magnon absorption sidebands in optical experiments \cite{Sell1967, Freeman1967, Freeman1968, Meltzer1969,Lohr1972,Tanabe1968}.} Further theoretical investigations are required to clarify the essence of this exciton delocalization and its coupling to electronic degrees of freedom through, e.g., dynamical mean field theory \cite{Wang2018}. Nonetheless, the mode-resolved and $\Delta$-dependent dispersive behavior presented here provides key constraints for reaching a consistent microscopic description of excitonic dispersion in charge transfer insulators.

%Figure 5

\section{Discussion}

Our results reveal the momentum-dependence of spin-singlet $dd$ excitations in NiX$_2$ compounds and we propose a ligand-mediated delocalization mechanism analogous to superexchange. Specifically, the microscopic interactions that give rise to the ${}^1A_{1g}$/${}^1E_g$ excitations are rooted in the $3d^8$ electronic configuration in octahedral symmetry, i.e. $dd$ excitations. Meanwhile, increasing metal-ligand charge transfer induces two intertwined, but distinct, effects. First, it renormalizes the intra-atomic Coulomb interactions at the Ni site and induces a corresponding reduction of the fundamental multiplet energies (Figs. \ref{fig:fig1}, \ref{fig:fig2}). Second, the excitations simultaneously develop an increasingly delocalized and propagating nature, independent of the magnetic phase (Figs. \ref{fig:fig3}-\ref{fig:fig5}). {These conclusions provide a self-consistent and comprehensive picture of the influence of metal-ligand charge transfer on the properties of multiplet excitations in charge-transfer insulators. }

The importance of measuring the exciton dispersion for unraveling its underlying nature has been stressed in several different contexts, including the alkali halides \cite{Abbamonte2008}, spin-state excitations in cobaltites \cite{Wang2018}, fractionalization in low-dimensional cuprates \cite{Schlappa2012, Martinelli2023, Bisogni2015}, molecular excitons \cite{Cudazzo2015, Cudazzo2013, Schuster2007, Yang2007} and the description of exciton-magnon sidebands in optics \cite{Sell1967,Freeman1967,Freeman1968, Kozielski1972, Day1976, Meltzer1969, Giordano1978, Pollini1980, Benedek1979}. For the latter, the momentum-dependence of the exciton state contributes to the sideband structure and also determines the mechanisms of intersite exciton-magnon coupling \cite{Sell1967, Freeman1967, Agranovich1968}, which are of strong relevance for interpreting photo-induced magnetic responses in transition-metal/vdW materials \cite{Sell1967, Freeman1967, Pogrebna2018, Genkin1982, Bae2022,Zhang2022, Ergecen2022,Matthiesen2023,Belvin2021,Hwangbo2021, Wu2019a}. While the momentum-dependence of spin-flip excitons has been indirectly inferred in optical experiments, our results provide the first direct evidence of their dispersive behavior. We note that the observed multiplet dispersion is distinct from the the case of fractionalized orbital excitations (orbitons) observed in cuprate spin chains \cite{Schlappa2012, Bisogni2015}, which is an emergent effect from low dimensionality. Instead, we expect the dispersion of the spin-flip excitons observed here to be a general phenomenon for the electronic excitations of charge-transfer insulators.

{More broadly, our temperature-dependent RIXS results (Fig. \ref{fig:fig4}) provide important context for the observation of these spin-flip $dd$ excitations in the optical regime \cite{Day1976, Robbins1976, Banda1986, Kuindersma1981a, Son2022, Kang2020a, Jana2023}. These multiplet transitions are optically dipole- and spin-forbidden, and therefore sensitive to the lowering of symmetry across (magnetic) phase transitions that are typically inferred from optical side-bands of bosonic origin (e.g. phonon, magnon \cite{Kozielski1972, Robbins1976, Giordano1978, Benedek1979, Pollini1980}). The complexity of such rich side-band structure, as well as the large energetic reormalization of the spin-forbidden peaks as a function of charge-transfer gap, has precluded consistent peak assignments and interpretations in the optical literature which we conclusively resolve. One consequence of the coupling to bosons is that the optical response of these exciton side-bands can be sensitive to the coherence of the magnon excitations \cite{Giordano1978, Pollini1980, Benedek1979}. The proposed effects of magnetic coherence are then likely attributed to the (magnetic) ${}^3A_{2g}$ ground state rather than the (non-magnetic) ${}^1A_{1g}$/${}^1E_g$ excited states. Conversely, the fundamental spin-flip multiplets, generally a high cross-section and direct RIXS process at the transition-metal $L$-edges, are well-defined and independent of magnetic order with lineshape limited only by a temperature-dependent Franck-Condon phonon broadening [Fig. \ref{fig:fig4}(c-f)].}

Finally, we have demonstrated several key design principles for tuning the functional exciton properties. We note that only the isolated ${}^1A_{1g}$ peak in NiBr$_2$ has a resolution-limited behavior at the lowest measured temperature ($T = 30$ K). This contrasts with the ${}^1A_{1g}$ excitations of NiCl$_2$ and NiI$_2$ which are broader than the experimental resolution and are partially/fully overlapped with multiplets. We hypothesize that a spin-orbit coupling induced hybridization of closely lying multiplets with distinct orbital configurations (e.g. $t_{2g}^6e_g^2$ and $t_{2g}^5e_g^3$) may be an important aspect limiting the intrinsic (low-$T$) linewidths. Furthermore, for making these modes optically bright with large oscillator strength, details of multiplet level sequencing (particularly the lowest excited state) and relative energetic proximity of different multiplet terms are known to be essential through, e.g., inter-system crossing and intensity borrowing mechanisms \cite{Kitzmann2022}. These are actively employed in the ligand-field engineering of spin-flip luminescence transitions in molecular systems \cite{Bayliss2020, Wojnar2020, Kitzmann2022}, which are direct molecular analogs to the spin-flip multiplets elaborated here. In this work, we have shown how the fundamental multiplet energies and their sequencing can be tuned by ligand field engineering and the charge transfer gap in the solid state. These underlying design principles could be fruitful for bringing on-demand and deterministic optical properties into the field of vdW materials.

\section{Conclusions}
In conclusion, we have extensively investigated the properties of the sharp, spin-singlet multiplet excitations of the nickel dihalides (NiX$_2$) using RIXS. We {demonstrated that} nearly resolution-limited $dd$ excitations are ubiquitous features of octahedrally-coordinated Ni$^{2+}$, which can be systematically tuned by the ligand and charge transfer gap. We have further established the roles of the self-doped ligand hole/charge transfer states and magnetism, ruling out a Zhang-Rice mechanism and revealing that the fundamental multiplet peaks are independent from long-range magnetic order. Most importantly, we { provide the first direct experimental evidence demonstrating that} these excitations are dispersive. We connected this behavior to an emergent effect of the increased self-doped ligand hole character of these excitations upon reduction of the charge transfer gap. Finally, we identified a potential mechanism for this exciton delocalization that is mediated by the ligand states in analogy to superexchange. Our RIXS results thus firmly establish the microscopic nature of these exciton states and provide a fundamentally distinct approach for tailoring collective electronic excitations in charge transfer insulators through their momentum dispersion.

{\it Acknowledgements}. We acknowledge insightful discussions with Maurits Haverkort and Ru-Pan Wang. This work was supported by the U.S. Department of Energy, Office of Science National Quantum Information Science Research Center's Co-design Center for Quantum Advantage (C2QA) under contract number DE-SC0012704 (X-ray spectroscopy measurements and data analysis), and by the US Department of Energy, BES under Award No. DE-SC0019126 (sample synthesis and characterization). This work was supported by the Laboratory Directed Research and Development project of Brookhaven National Laboratory No. 21-037. This research used beamline 2-ID of the National Synchrotron Light Source II, a U.S. Department of Energy (DOE) Office of Science User Facility operated for the DOE Office of Science by Brookhaven National Laboratory under Contract No. DE-SC0012704. This work was supported by the U.S. Department of Energy (DOE) Office of Science, Early Career Research Program. C.A.O. and Y. T. contributed equally to this work.

\appendix

\section{Methods}

\subsection*{Sample growth and preparation}
All samples were prepared using chemical vapor transport (CVT). NiCl$_2$ was synthesized using stoichiometric ratios of Nickel powder (Sigma Aldrich, 99.9\%) and TeCl$_4$ (Sigma Aldrich, 99.8 \%), at a temperature gradient of 760 $^\circ$C to 730 $^\circ$C for 72 hours before being cooled naturally to ambient conditions. The temperature ramp-up time was 72 hrs. Single-crystal NiBr$_2$ was grown from NiBr$_2$ powder (anhydrous, $>99.9$\%, Sigma-Aldrich), at a temperature gradient 650 $^\circ$C to 600 $^\circ$C. Single-crystal NiI$_2$ was grown from elemental precursors with molar ratio Ni:I = 1:2, at a temperature gradient 700$^\circ$C to 500$^\circ$C as described previously \cite{Song2022}. The magnetic susceptibility was measured using a magnetic property measurement system (MPMS-3, Quantum Design Inc.) for NiI$_2$/NiBr$_2$ and a physical property measurement system (PPMS, Quantum Design Inc.) using the vibrating sample magnetometer (VSM) option for NiCl$_2$. The magnetic susceptibility of the bulk crystals confirm magnetic transitions at $T_N = 53$ K for NiCl$_2$, $T_{N,1} = 45 $ K and $T_{N,2} = 22$ K for NiBr$_2$ and $T_{N,1} = 75$ K and $T_{N,2} = 60$ K for NiI$_2$. Magnetic susceptibility data are shown in the Supplemental Material for NiCl$_2$ and NiBr$_2$ and in Ref. \cite{Song2022} for the NiI$_2$.

The samples were aligned using a Bruker-GAADS Co-$K_\alpha$ ($\lambda = 1.7902$ \AA) x-ray diffractometer to place the $\mathbf{a}^*$ direction in the scattering plane for RIXS experiments. The lattice parameters were determined to be $a = 3.465(12)$  \AA ~and  $c = 17.304(46)$ \AA ~ for NiCl$_2$, $a = 3.648(13)$ \AA~and $c = 18.412(52)$ \AA ~ for NiBr$_2$ and $a = 3.934(15)$ \AA ~and $c = 19.809(61)$ \AA ~ for NiI$_2$ which were determined by single crystal diffraction from the (006) and (104) reflections. Samples are aligned in air, cleaved in a high-purity nitrogen-filled glovebox (H$_2$O and O$_2$ $< 0.1$ ppm) and stored in vacuum for transport to the X-ray beamline. For NiI$_2$, we left the as-grown surface uncleaved for XRD alignment, with air exposure of approximately 15 minutes. Cleaving of the as-grown surface inside a nitrogen-filled glovebox after alignment revealed protected surfaces without visible degradation. The sharp multiplet features in agreement with optical spectra \cite{Kozielski1972, Banda1986, Kuindersma1981a} and low diffuse scattering of the soft X-ray beam confirms high-quality samples and flat vdW surfaces for all samples. 

\subsection*{X-ray absorption and resonant inelastic X-ray scattering experiments}
XAS and RIXS measurements at the Ni $\mathrm{L}_3$-edge (852 eV) were carried out at the 2-ID SIX beamline at the National Synchrotron Light Source II, Brookhaven National Laboratory. $\sigma$ polarization was applied for the incident X-rays for all measurements. XAS was recorded in total fluorescence yield (TFY) using a photodiode inside the soft X-ray chamber. RIXS spectra were recorded with high-resolution of $\Delta E = 31$ meV for all measurements. The sample temperature was kept at 40 K unless specified. Laboratory-prepared and sealed samples were transferred from vacuum into the ultra-high vacuum (UHV) loadlock of the X-ray chamber with minimal air exposure and kept under UHV conditions for the duration of X-ray experiments. NiI$_2$ samples are more hygroscopic, and were loaded into the vacuum chamber within a high-purity nitrogen environment. 

\subsection*{Charge-transfer multiplet calculations}
We performed charge transfer multiplet (CTM) and crystal field multiplet (CFM) calculations using the Quanty software \cite{Haverkort2014,Haverkort2012,Lu2014}. For the CFM calculations, we consider the Ni-$3d$ orbitals in the basis set with an octahedral crystal field ($O_h$ CF). For CTM calculations, the symmetrized $X = $ Cl/Br/I molecular orbitals of $t_{2g}$ and $e_g$ symmetry are explicitly included. For the main text, we restrict all calculation to $O_h$ symmetry, while the effects of the trigonal $D_{3d}$ distortion are considered in the Supplemental Material. Core-level spectra for the Ni-$L_{2/3}$ edges are calculated by considering $2p\to3d$ dipole transitions using the Green's function formalism \cite{Haverkort2014,Haverkort2012,Lu2014}. All spectra are calculated in the experimental polarization conditions. 

The parameters for the multiplet calculations include the Coulomb interactions at the Nickel site, parameterized as the direct Slater integrals $F^2_{dd}$/$F^4_{dd}$ in the initial/final RIXS state $(3d^8)$ and by the direct integrals $F^2_{dd}$/$F^4_{dd}$/$F^2_{pd}$ and exchange integrals $G^1_{pd}$/$G^3_{pd}$ in the intermediate RIXS state $(2p^53d^9)$. The atomic spin-orbit coupling (SOC) in the $3d$ and $2p$ nickel states are also included. The Slater integrals and SOC parameters for each electronic configuration are taken from Hartree-Fock values tabulated by Haverkort \cite{Haverkort2005}. For CTM calculations, all Slater integrals are uniformly scaled to atomic values (80\% of HF values). Additional parameters include the Coulomb repulsion parameters $U_{dd}$/$U_{pd}$ = 5.0/7.0 eV, respectively (taken from photoemission experiments \cite{Zaanen1986, VanDerLaan1986}), the charge transfer energy $\Delta$ and the metal-ligand hybridization $V(e_g)/V(t_{2g})$. We use the empirical relation $V(t_{2g}) = 3/5 \times V(e_g)$ and fit the XAS/RIXS spectra using $V(e_g)$/$\Delta$. Full details of the calculations can be found in the Supplemental Material.

%\bibliography{NiX2_Bib_20230711_Corrected}


\end{document}